\documentclass[conference]{IEEEtran}
\IEEEoverridecommandlockouts
\usepackage{cite}
\usepackage{amsmath,amssymb,amsfonts}
\usepackage{algorithmic}
\usepackage{physics}
\usepackage{graphicx}
\usepackage{textcomp}
\usepackage{xcolor}
\usepackage{titlesec}
\def\BibTeX{{\rm B\kern-.05em{\sc i\kern-.025em b}\kern-.08em
    T\kern-.1667em\lower.7ex\hbox{E}\kern-.125emX}}
\titlespacing*{\subsection}
{0pt}{3.0ex plus 1.0ex minus .2ex}{2.3ex plus .2ex}

\begin{document}

\title{From Problem to Solution: A general Pipeline to Solve Optimisation Problems on Quantum Hardware
}

\author{
\IEEEauthorblockN{%
Tobias Rohe\IEEEauthorrefmark{1},
Simon Grätz\IEEEauthorrefmark{1},
Michael Kölle\IEEEauthorrefmark{1},
Sebastian Zielinski\IEEEauthorrefmark{1},
Jonas Stein\IEEEauthorrefmark{1}\IEEEauthorrefmark{2}
and Claudia Linnhoff-Popien\IEEEauthorrefmark{1}}
\IEEEauthorblockA{\IEEEauthorrefmark{1}\textit{LMU Munich, Institute for Computer Science, Munich, Germany}}
\IEEEauthorblockA{\IEEEauthorrefmark{2}\textit{Aqarios GmbH, Munich, Germany}}
\IEEEauthorblockA{\{tobias.rohe, michael.koelle, sebastian.zielinski, jonas.stein, linnhoff\}@ifi.lmu.de}
}

\maketitle

\begin{abstract}
With constant improvements of quantum hardware and quantum algorithms, quantum advantage comes within reach. Parallel to the development of the computer at the end of the twentieth century, quantum software development will now also rapidly gain in importance and scale. On account of the inherent complexity and novelty of quantum computing (QC), as well as the expected lack of expertise of many of the stakeholders involved in its development, QC software development projects are exposed to the risk of being conducted in a crowded and unstructured way, lacking clear guidance and understanding. This paper presents a comprehensive quantum optimisation development pipeline, novel in its depth of 22 activities across multiple stages, coupled with project management insights, uniquely targeted to the late noisy intermediate-scale quantum (NISQ) \cite{preskill2018quantum} and early post-NISQ eras. We have extensively screened literature and use-cases, interviewed experts, and brought in our own expertise to develop this general quantum pipeline. The proposed solution pipeline is divided into five stages: Use-case Identification, Solution Draft, Pre-Processing, Execution and Post-Processing. Additionally, the pipeline contains two review points to address the project management view, the inherent risk of the project and the current technical maturity of QC technology. This work is intended as an orientation aid for all stakeholders involved in the development of QC applications and should therefore increase the chances of success of quantum software projects. We encourage researchers to adapt and extend the model where appropriate, as technological development also continues. 
\end{abstract}


\section{Introduction}
Quantum Computing (QC) is a rapidly evolving field with great efforts towards new hardware and potentially disruptive algorithms. Current research is primarily concentrated on the basic principles of QC hardware and algorithms, aiming to achieve quantum advantage. However, it's equally important to emphasise the development of processes for creating production-ready quantum software. In this context, the procedures and key elements of quantum software development become crucial. \\
Through the inherent differences in principles of classical and quantum computing \cite{gill2022quantum}, different structures of quantum algorithms, and therefore also of quantum software and its development arise \cite{zhao2020quantum}. Particularly at the beginning, where QC is in its early years, the restricted capabilities of hardware-algorithm combinations have to be understood and considered, as only very few use-cases are suitable for the application of QC. To make matters worse, in these early years, we have observed that non-QC experts often have a greater lack of understanding of QC technologies, their capabilities and their applications. As research has shown, in classical software development it is of great importance to build a solid understanding of the developing-process and mutual understanding of expectations, expertise's, interdependencies and the task itself across the team \cite{curtis1988field, rose2007management, moe2008understanding, cha2019prioritising}. We do not expect this to be any different in the development of QC software solutions.\\
We have identified this lack of structure regarding the quantum software developing process in terms of clearly defined development phases and the activities carried out in these phases. Particularly in times of noisy intermediate-scale quantum (NISQ), where hardware is limited \cite{preskill2018quantum} and QC is only appropriate for a niche of problems, the development process is challenging and yet not outlined enough. We want to close this gap with a pipeline that guides the development of quantum solutions from its initial problem, to a ready to implement solution. \\
With our proposed pipeline we target the late NISQ and early post-NISQ time, focusing on optimisation use-cases and universal QC, while trying to keep the pipeline easy to adapt to a general context and other areas like quantum machine learning applications. Therefore, focusing on solutions tailored for distinct and specific instances of problem categories rather than for seeking generic, one-size-fits-all solutions. The model is tailored to a project-based setting, where an institution or consortium has identified a problem they want to solve with QC technology. We do not solely focus on technical details, but also try to include project management aspects, i.e. to consider all stakeholders. The depth of the $22$ activities described, the inclusion of project management aspects, the focus on quantum optimisation solutions, and the focus on late NISQ and early post-NISQ era make the work novel.\\
Our pipeline can help project teams to streamline their process and facilitate communication. This applies for research teams in academic institutes and companies, business consortia, management teams, as well as for students and early adopters who want to experiment with this new technology. We hope to lower entry barriers, increase project success and reduce the risk of project failure with this proposed pipeline. All with the common goal in mind to bring QC into application and ready to use for the future practitioners. \\
This paper is structured as follows: Chapter $2$, \textit{Related Work}, delves into the literature that forms the foundation of our study. In Chapter $3$, we describe our research methodology and the process through which we developed the current model. Chapter $4$ presents our proposed pipeline in comprehensive detail. Finally, in Chapter $5$, we draw conclusions from our work. 

\section{Related Work}
Developing quantum software solutions and integrating them into existing workflows, IT infrastructures, and established processes presents a substantial challenge due to the distinctive characteristics and the rapid development of quantum technologies. In this section, we explore existing literature within the realm of quantum software development, implementation, quantum workflows (QWFs), as well as  solution paths for distinct use-cases. We argue that literature from all these research fields have contributed to our work, paving the way to a general quantum development pipeline.

Close to hardware, Leymann and Barzen ($2020$)~\cite{leymann2020bitter} delve into factors influencing quantum algorithm implementations, focusing on pre-processing, state preparation, unitary transformation, measurement, and post-processing. The authors argue that too little attention is paid to these phases together, even though they have a major impact on the implementation on NISQ devices. According to the authors, the pure focus on the unitary transformation falls in general short to assess the successfully execution of algorithms here. Leymann and Barzen provide the details for the adaptations that need to be made to algorithms in order to match the corresponding NISQ devices. 

Another study by Leymann et al. ($2020$)~\cite{leymann2020quantum} explores QC in cloud environments, emphasising the technical realisation of hybrid workflows and algorithms for cloud environments. Particular attention is paid to the fit between the quantum algorithm and the hardware stack used, whereby a viable architecture for such an integration has been developed. A suitable architecture, like the one proposed in their work, plays a key role in the development of a quantum solution, as it determines significant design decisions on the software side.  

From the field of QWFs, in $2020$ the Quantum Modeling Extension (QuantME) package expands existing (classical) workflow languages for the adaptation and modelling of QWFs~\cite{weder2020integrating}. It primarily focuses on the description and representation of automatically executable QWFs, extending existing frameworks, and assessing their portability and applicability. The researchers work along the technical structure of a quantum solution, describing essential parts of these solutions. 

In their $2022$ paper, Ahmad et al. explore the field of Quantum Software Engineering, focusing specifically on its architecting processes and the associated human roles and tools~\cite{ahmad2022towards}. The paper presents a detailed examination of five key architectural activities: Requirements, Modeling, Implementation, Validation, and Deployment. These activities are exemplified through a case study on quantum key distribution, which serves to practically demonstrate the theoretical concepts discussed. Furthermore, the authors identify four critical human roles within the quantum software engineering domain: Quantum Domain Engineer, Quantum Software Designer/Architect, Quantum Code Developer, and Quantum Simulation Engineer, providing insights into their responsibilities and the tools they can utilise. Although the research presented by Ahmad et al. aligns closely with our own work, our study diverges by simplifying the human roles involved and instead expanding the development pipeline. Our approach introduces numerous additional activities across a five-phase model, thereby enriching the depth and scope of the development pipeline itself.

Quetschlich et al. ($2023$)~\cite{quetschlich2023towards} proposes a framework consisting of four quantum steps towards a QC solution: algorithm selection, problem encoding, execution on a quantum device, and solution decoding. The researchers goal is thereby to propose a framework which is capable for a, yet to realise, automated QC solution generation, to shield the QWF steps from the user. Their framework defines the required input and output specifications for such an automation. While this framework is again close to architecture and infrastructure decisions, our work will focus more on the steps in today's and near-term future development of quantum solutions, which are still made in a manual fashion, as the authors also agree on.

Going away from an architecture, integration and workflow perspective, Luckow et al. ($2021$)~\cite{luckow2021quantum} follows an application oriented approach, presenting several quantum use-cases which are of high-value for the automotive industry. The paper does thereby offer a problem description, names the problem domain, problem class, mathematical formulation, and algorithmic solution of these use-cases. Without going into technical details, the authors outline on a high-level the proposed use-cases with possible solution paths in the quantum realm. Their work provides insights into quantum solutions, which we further leverage and extend in our framework. 

A comprehensive survey on quantum software engineering in general is presented by Zhao et al. ($2020$)~\cite{zhao2020quantum}. Their research covers the various phases of the quantum software life cycle, including requirement analysis, design, implementation, testing, and maintenance. The elements listed here can also be found in parts in our pipeline. The authors outline the parallels to classical software development cycles, while at the same time they adapt the model, where necessary, to the specific characteristics of QC. As we will later see, our pipeline also picks up inspiration from related technological fields like artificial intelligence and their development pipelines (see~\cite{gabor2020holy}). 

Similar to Zhao et al., Weder et al. ($2022$)~\cite{weder2022quantum} present a high level quantum software development lifecycle, drawing also parallels to classical software engineering practices. The researchers define key phases such as requirement analysis, architecture \& design, testing, deployment, observability, and analysis, providing a holistic perspective on quantum software development. Both works are thereby outlining a very similar lifecycle, describing a similar process. 

A systematic literature review on software architectures for quantum computing systems is provided by Khan et al. ($2023$), offering a comprehensive overview of the current research in the field~\cite{khan2023software}.

The aforementioned literature collectively form the foundation of existing research in our field. However, there remains a research gap in today's and near-term future best practice for developing solutions for optimisation problems using QC. Our work aims to address this gap by providing a structured and comprehensive pipeline to develop quantum solutions for optimisation use-cases on a project to project level. We are not focusing on architecture, infrastructure and the software life-cycle, but rather on decision management, collaboration, and development steps. Our pipeline does provide stages and decision, but also comprise more technical sub-items, which may enhance the quality of the final solution. In contrast to other work, we are targeting the late NISQ or post-NISQ period, whereby the quantum advantage for specific problems has already been achieved. 

\section{Method \& Proceeding}
Based on the examined literature, we conducted an in-depth analysis to discern recurring patterns in quantum software development for optimisation problems. Even though the papers covered above have different scopes, ranging from technical workflow concepts to more conceptual development aspects, we observe that specific patterns and processing steps reappeared throughout the literature, albeit often under distinct nomenclature and with nuanced variations. Nevertheless, the core structure and functionality of these patterns exhibited only minor deviations. We thus extracted these core patterns and organised them into a hierarchical framework. This effort yielded an initial iteration of our quantum development pipeline, comprising five major processing stages. Throughout our investigation, we continuously refined our pipeline model and incorporated new activities as necessary. We will go into more detail regarding the individual elements and their purpose in the next chapter. 
After developing an initial pipeline, we have verified its applicability and completeness. For this purpose, we scrutinised scientific literature about implemented use-cases \cite{chai2023towards, amaro2022case, nannicini2019performance, mugel2022dynamic, stollenwerk2020toward} and compared their methodology with our pipeline. If a processing step was identified that is relevant and was not yet included in our pipeline, this aspect has now been integrated. \\
After theoretical verification we added a final phase of practical verification, where we approached a quantum software development company to verify the applicability of the developed pipeline. We presented our pipeline to several of their quantum software engineers and discussed the individual phases and activities with them. We were able to incorporate valuable feedback from their practical experiences through these interviews and adapt the pipeline to the current workflows and the anticipated changes to the late NISQ and early post-NISQ era. Drawing upon them, we resolved lingering questions and completed the refinement of our model. Due to current hardware restrictions, it was not possible to implement and use the pipeline in practice, especially with regard to its focus on the late NISQ and the early post-NISQ era. The following section will now present the final results of this development process.

\section{Quantum Pipeline}
\begin{figure*}[!t]
  \centering
  \includegraphics[width=\textwidth]{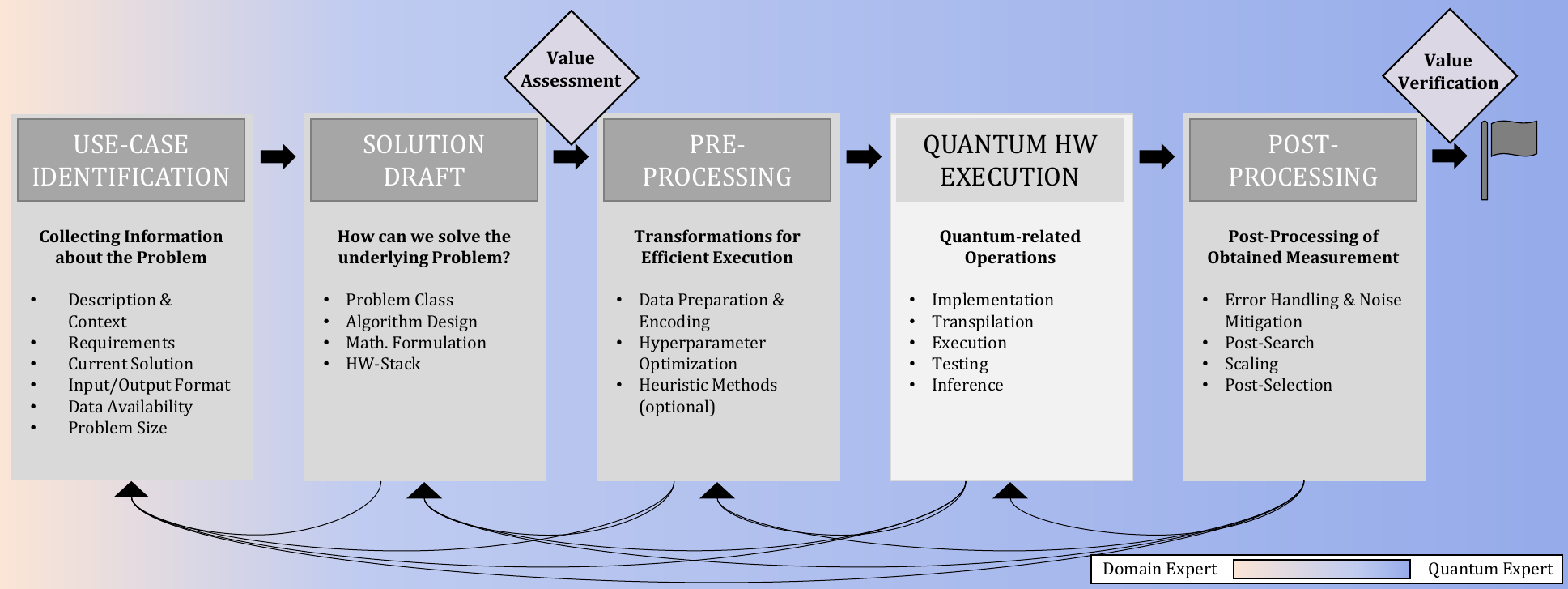}
  \caption{A General Quantum Development Pipeline. The pipeline is organised in stages, depicted as boxes, activities, listed within the boxes, and review-points, depicted as rhombi. It is traversed from left to right, with the review points placed between the stages. It is also permitted to iterate back through the pipeline depending on the circumstances, as this can never be completely avoided in practice. While the domain expert plays a pivotal role in the ``Use-case Identification'' stage at the beginning, the role of the quantum expert becomes increasingly important as the pipeline execution progresses. For the review points ``Value Assessment'' and ``Value Verification'' both experts are again fully needed.}
  \label{fig:pipeline}
\end{figure*}

In this chapter, we will introduce the quantum development pipeline and explain the purpose of each of the stages, activities and review points our pipeline encompasses. An overview of the pipeline is shown in Figure \ref{fig:pipeline}. 

Our pipeline consists of 5 stages: Use-case Identification, Solution Draft, Pre-Processing, Quantum Hardware-Execution, and Post-Processing. A \textit{stage} in this context is defined as a distinct phase within the overall pipeline. These stages are designed to be transitioned throughout the project, from its inception to the final quantum solution. Each stage does thereby consists of multiple activities. An \textit{activity} is a specific task or operation within a stage, contributing directly to the achievement of that stage's objectives. As we are talking about a highly complex technology which is just emerging, new insights during the project, in particular later stages, can cause consequences for supposedly completed stages and activities. It is therefore possible to jump back to these respective stages and activities, incorporating new insights. In addition, activities often have an optional character and are not appropriate for every problem. Individual activities can therefore be omitted or only partially fulfilled, without jeopardising the solution. The included \textit{review points}, ``Value Assessment'' after the second stage, and ``Value Verification'' at the end of the pipeline, question whether a continuation of the solution development makes under technical and economic viewpoints sense, respectively whether the project was a success and should be further continued and deployed.

\subsection{Use-case Identification}
The initial stage of our quantum development pipeline is the use-case identification stage. Its primary objective is to attain a comprehensive understanding of the problem to be solved among all stakeholders, which is a crucial aspect to meet customer and end-user expectations. Beside facilitating meeting expectations, the stage should also reduce the number of requirement changes throughout the project. Research indicates that insights into the underlying problem and objectives of the software development process, when gained late, correspondingly increase the need for design refinements \cite{nurmuliani2004analysis}. Those refinements are new, modified or deleted requirements, that specify, for example, the functionality of the yet to develop software. The stage of use-case identification therefore has qualitative character and sets the basis for the subsequent stages and reviews. Our use-case identification stage encompasses six key activities, each contributing to a thorough understanding of the problem:

\paragraph{\textbf{Description \& Context}} 
The first activity demands for a detailed description and context, in classical software development, also called environment, in which the underlying problem is embedded. Use-case relevant information should be gathered here to provide all stakeholder a common starting point and a mutual understanding, laying the groundwork for a successful transition through the pipeline. Academic work in classical software development has identified four major categories of environmental factors: project, team, external stakeholders, and organisation \cite{xu2007software}. These categories contain factors like project size, budget, stakeholders' background, process maturity, senior management commitment, and many more that also apply to quantum software development. All twenty factors listed in the paper, as well as other questions arising from the nature of the matter, help to understand the strategic position of the customer and the end-user, the impact of the problem on the entire value chain, its future relevance and its impact on other IT related areas. The goal is to enable developers to fully understand the customer and end-user situation and their motivation to develop the use-case. This in turn has a significant influence on tailoring the software process \cite{xu2007software}. Furthermore, the decision quality at the review points get enhanced, which we will discuss further later on.

\paragraph{\textbf{Requirements}} 
This activity goes beyond the initial understanding of the use-case, capturing requirements like required execution times, budget-constraints, or integration obstacles \cite{sommerville1997requirements}. Also the optimisation objective(s) and constraints become identified here. The activity tries to ensure that all mission critical aspects of the postulated solution get considered and formulated. The requirement analysis for quantum software is thereby very similar to that of classical software.  

\paragraph{\textbf{Current Solution}} 
The current solution activity identifies the approach with which the problem is handled right now. Although sometimes the problem at hand does not get processed in a specific manner yet, sometimes classical heuristics or approximation algorithms are in place. These existing solutions can be a further source to better understand the problem and assist in the later following value verification review, serving as benchmark to compare to.  

\paragraph{\textbf{Input/Output Format}} 
The input / output format activity asks for details about the input format which is fed into the later developed quantum solution, building its starting point. In our research and hands-on experience we have seen, that even for similar problems, different data input formats are possible. A different input format can thereby have consequences for pre-processing steps, namely the data preparation activity, but potentially also for the later following algorithm selection. The required output format, in contrast, is part of the endpoint of the quantum solution yet to be developed. It is the final solution the post-processing stage should output for the given optimisation problem at hand. Through the post-processing stage the output-format of the algorithm can get transformed, if possible, to the finally required format. In the simplest cases, for example for the maxut problem, this can be a binary string.

\paragraph{\textbf{Data Availability}} 
Assessing data availability is in two ways a not to underestimate activity in the development process of quantum solutions. Firstly, it involves gathering historical data sets of past problem instances, important not only for training and testing quantum algorithms during the development phase but also fundamental for benchmarking various solutions at later stages. Secondly, the analysis of historical problem instances can provide insight into the typical structure and complexity of the expected problem instances, enhancing our understanding. The goal is to identify patterns in solution variability and similarities across the instances, which can be exploited to enhance solution efficiency. Here, some warm start techniques and (classical) heuristic methods are particularly suited for similar problem instances \cite{poloczek2016warm}.

\paragraph{\textbf{Problem Size}} 
The third data-related activity asks about the expected size of the problem instances to solve. Often, we face varying problem sizes across problem instances, which further complicates the development of an appropriate solution. The size of problem instances significantly impacts the determination of suitable solutions, the decision to employ heuristic approaches, and the selection of hardware and algorithms. Furthermore, the problem size is a major determinant of the feasibility to find the global optimum of the optimisation problem, particularly in the late NISQ / early post-NISQ era. We will further discuss this issue at the first review-point. A deep understanding of the problem size and its variance across problem instances is indispensable, having also consequences on the problem encoding decision.\\

The above stated activities require a high level of communication and coordination across the different stakeholders of the project. In particular, the use-case provider and the end-user have to explain the context of the underlying problem. However, developers need to get a sound understanding of the problem, not only by reading the provided material, but much more by actively asking and questioning given answers. A misunderstanding and / or sloppy conducted use-case identification stage can, and most probably will, lead to an unsatisfying outcome, as it is also the case in classical software development \cite{williams2006change, chari2018impact}. However, even with great efforts it will be difficult to communicate all relevant information at this early stage. Use-case providing actors often discover new aspects of the problem during the pipeline execution, to which they were not aware of before, making iterations necessary. Therefore, this stage is never fully terminated, but will regularly be updated during the development process. After all stakeholders gained a well-grounded understanding of the use-case, the next stage ``Solution Draft'' can be initiated.

\subsection{Solution Draft}
Once the use-case has been identified and described, a suitable solution draft must be formulated. In order to achieve such a draft, we have outlined four activities that will lead the developer to achieve this objective.  

\paragraph{\textbf{Problem Class}}
Based on the problem description, a distinctive problem class, or rather a combination of problem classes, such as flow-shop scheduling, vehicle routing, or knapsack problem, can often be identified. An assignment can help developers to orient themselves alongside existing solutions for similar use-cases. The inherent structure of these specific problem classes, as well as already conducted research and development can help to simplify the ongoing development process. However, it needs to be mentioned that sometimes such a clear assignment is not feasible, as certain problems may present unique or novel challenges that do not align neatly with established problem categories (for example incorporating time constrains into the capacitated vehicle routing problem (CVRP) within the realm of well-studied 21 Karp's problems \cite{karp1972reducibility}). 

\paragraph{\textbf{Algorithm Design}}
We first choose a suitable algorithm for the underlying problem before we create the necessary mathematical formulation. This activity goes beyond a simple algorithm selection, as every major (hybrid) quantum algorithm has several sub-variants and configurations, suitable for different problem-settings. For example the Variational Quantum Eigensolver (VQE): The selection of an appropriate Ansatz \cite{sim2019expressibility}, training variant \cite{liu2022layer, stein2023introducing}, optimiser \cite{bonet2023performance} and measurement strategy \cite{tang2021qubit}, should be outlined upfront. For sure, changes of the algorithm design can also be conducted in a later stage of our developing pipeline, as some decisions might also be done by a trial-and-error approach or depend on decisions made later. An illustrative example here is the generation of a hardware-efficient Ansatz for the VQE algorithm, which of course depends on the selected hardware. However, the inherent algorithmic structure should already be determined here, even if it may have to be adapted again later. It is advisable, as it is also usual in the development of classic solutions, to design and implement several algorithms in parallel in order to compare them with each other at the end of the pipeline. This is particularly relevant for algorithms for similar problems where no clear performance winner has yet been determined, e.g. the quantum approximate optimisation algorithm (QAOA) and VQE algorithm. 

\paragraph{\textbf{Mathematical Formulation}}
The mathematical formulation activity involves formulating the problem as a mathematical model or transforming an existing model into a model suitable for quantum algorithms, e.g. into quadratic unconstrained binary optimisation (QUBO) problems or polynomial unconstrained
optimization (PUBO) problems. Therefore, the variables of the optimisation problem, all relevant constraints, as well as the final cost function have to be formulated in a mathematical precise form. Here, constraints can often be further broken down into hard and soft constraints, therefore, constraints which must not be broken, respectively should not be broken. Additionally, also the objective function is often multi-dimensional, having a main objective, like cost minimisation or revenue maximisation, as well as side objectives, like the reduction of CO2 emissions or lead times. The balancing of the different optimisation objectives can thereby also be adapted during the development process. The mathematical problem formulation~\cite{stein2023evidence} as well as the formulation of the cost function~\cite{cerezo2021cost} can significantly affect the performance of the quantum algorithm and should therefore be chosen with care.

\paragraph{\textbf{Hardware-Stack}}
Finally, after formulating the problem and designing the proposed algorithm, a suitable hardware-stack must be chosen to fit the previously defined pieces. The hardware-selection does not only depend on the number of qubits a device has, but also on quality characteristics like gate execution times, noise levels, read-out errors, etc., as well as other non-qualitative aspects like availability and costs. Additionally, the hardware-stack contains further elements like compiler or native gate sets, which significantly impact the performance ultimately achieved \cite{bandic2022full}. This step is much more complicated in quantum computing compared to classical computing, as the hardware differs and varies on many more levels. Choosing the right, most efficient hardware stack is therefore a major challenge \cite{weder2021automated}.\\ 

The interdependencies between individual activity elements, as well as the interdependencies across the activities, make this stage in particularly challenging. Further research regarding well suited combinations of algorithmic- and hardware-elements is needed here.

\subsection{Value Assessment}
After completing the first two stages of our quantum development pipeline, the first review is conducted. The central question is whether the development should be continued, or the expected value of the project is too low to make the necessary investments to continue. This decision is driven by two main aspects: technical feasibility and strategic benefit. As the first two stages of the pipeline have been traversed, a well-founded initial understanding of the problem at hand and its characteristics should be achieved. Based on this understanding, quantum experts can assess whether the problem is in principle suitable to be solved with QC in the near future, or whether a successful application of QC technology is still many years away. While many current projects appear feasible on quantum computers only in the distant future, potentially decades away, major advancements in hardware development could shorten this timeline in the coming years. In addition, the strategic component of such projects also plays an important role here. \\
Even if the quantum advantage for first problems may be reached in the late NISQ / early post-NISQ era, it will not make sense for all problems to be solved on a quantum computer yet. Although, exploring quantum solutions  can also be useful for these problems from certain strategic points of view. The magnitude of expected disruption, or the resource availability of the development-financing stakeholder are only two aspects which need to be considered here. For companies in industries anticipating significant disruption from the successful application of QC, early investments in the development of quantum solutions might be legitimate, particularly for those with substantial financial resources. These considerations are inherently complex and warrant further scientific investigation, also from a business research perspective.

\subsection{Pre-Processing}
With the pre-processing stage, all relevant preliminaries for the later following implementation and execution should be answered and prepared. We will now outline three activities which can be conducted in order to reach the execution stage, keeping in mind that our model is still highly iterative. 

\paragraph{\textbf{Data Preparation \& Encoding}}
Similar to the field of artificial intelligence, data plays, as an input, a crucial role for optimisation problems. If the input-data is not adequately collected, pre-processed or interpreted, a wrong model of the problem instance might get generated \cite{mytkowicz2009producing}. A solution that appears perfect for the model may in fact be incorrect for the underlying problem to solve. Therefore, data integration, data cleaning, variable selection, and missing data handling are steps that should not be neglected in the data preparation step. Further, data normalisation and data transformation can help to enable a more efficient execution of the later following quantum algorithm. The activity of data preparation can be well oriented to existing literature in the field of artificial intelligence, with small adaptations to be made \cite{kotsiantis2006data, garcia2015data, fan2021review}. \\
One the other side, the task of data encoding represents one of the biggest differences between classical and quantum solutions for optimisation and machine learning problems. While data encoding / loading is also a major challenge for machine learning applications in classical computing ~\cite{zhang2016materialization, kumar2017data, chai2022data}, this problem is much more challenging in the realm of quantum computing. Due to the small number of qubits and the short coherence times, loading even very small amounts of data represents a major challenge ~\cite{preskill2018quantum, park2019circuit, leymann2020bitter}. This will continue to be a problem in the late NISQ and early post-NISQ era. The choice of a suitable data encoding, for example techniques like angle or amplitude encoding ~\cite{larose2020robust, weigold2021expanding}, as well as the use of yet-to-be-developed efficient QRAMs ~\cite{giovannetti2008quantum, park2019circuit, weigold2021expanding} is a major challenge in the development process of quantum algorithms, which does not necessarily appear in this form in classical computing. 

\paragraph{\textbf{Hyperparameter Optimisation}}
Quantum algorithms and corresponding mathematical models possess several hyperparameters that can significantly impact the solution quality. Hyperparameters in the QAOA or VQE algorithm include, for example, the number of layers used \cite{sim2019expressibility}, the optimiser configuration \cite{bonet2023performance}, and the training type \cite{liu2022layer}. In order to find optimal hyperparmeters, there are various strategies that can be applied. Grid Search, Random Search and Bayesian Optimisation are just some ways to improve hyperparameters and therefore also solution quality.  

\paragraph{\textbf{Heuristic Methods (optional)}}
Right away we want to mention the optional character of the heuristic methods activity. Without this activity the development aims for finding the global optimum for the given and modelled problem instance at hand. However, finding the global optimum is incredibly difficult (more formally spoken often NP-hard) \cite{karp1972reducibility}. Beside this intrinsic difficulty of the problem, formalised in the complexity class which is based on the scaling characteristic, the sheer size of the problem contributes to the practical insolubility. A reduction of problem size is therefore a valid strategy to make the problem better processable. Developers, with consultation of domain-experts, can apply heuristic methods which try to either decompose the problem into smaller sub-problems \cite{ralphs2003capacitated} or apply pruning strategies, which make assumptions and simplifications to reduce the problem size. While those methods can significantly reduce problem size, this comes at the cost of loosing information, and therefore often the possibility of finding the global optimum. By deciding to conduct such heuristic methods, the goal of the solution shifts to finding a (very) good solution, a solution better than the best classical approximation, instead of finding the globally optimal solution. Therefore, the decision must be well thought out and in line with the overall development objective.

\subsection{Quantum Hardware Execution}
To this point, no program code has been executed on the chosen quantum hardware. Based on experience and codified knowledge, rather than on experiments around the outlined quantum solution, a concept has been developed yet. The four activities, implementation, execution, testing, and inference are now about turning the collected ideas into code in order to find the most suitable solution. While theoretical deductions are important for determining an experimental starting point, the complexity, size and fast evolution of the field make predictions about the optimal solution path difficult. This stage is particularly iterative and interacts strongly with the previously conducted stages. Through new insights it may be ultimately necessary to reevaluate existing assumptions and to try out new approaches.

\paragraph{\textbf{Implementation}}
In this activity, previously conceptualised ideas and thoughts are transformed into executable code, thereby laying the groundwork for the subsequent execution activity. At this point, at the very latest, a specific hardware stack must be defined as targeted architecture, as otherwise hardware-software incompatibilities may arise. By implementing the quantum solution, developers might discover new aspects, characteristics, and opportunities to enhance the quantum solution. Such insights are of high value and should be gathered, analysed, and transferred back to the iterative approach outlined here. 

\paragraph{\textbf{Transpilation}}
The transpilation process is a critical activity, differing significantly from classical computing compilation. Quantum transpilation is designed to translate the inputted quantum circuits, for example the Ansatz used in the VQE algorithm, with regards to the practical limitations, namely the topology of the quantum processing unit (QPU) and its native gate set, into a circuit executable on the respective device~\cite{murali2019noise}. A transpiler integrates optimisation strategies to counter quantum decoherence and operational errors at the same time~\cite{wilson2020just}. The efficiency of the transpiler does thereby have a significant impact on the performance of the final application which should not be underestimated~\cite{liu2021relaxed}.

\paragraph{\textbf{Execution}}
Once the code has been realised, it can be executed on real quantum hardware to generate solutions for the considered problem. Most often, the code runs in a hybrid environment, an interplay of classical and quantum hardware. Here too, special requirements of hardware vendors and providers might need consideration. However, the design implementation is often compatible with a variety of quantum hardware architectures, each characterised by unique technological paradigms such as different physical methods of qubit realisation. For instance, the same algorithm design can typically be executed on superconductor-based, ion traps, or other systems with little or no modification. Depending on the resources and status of the project, an execution on different hardware-platforms should be considered, even if code adaptations are therefore necessary. 
We have deliberately not included the simulation of QC applications in the pipeline as a standalone activity. While such simulations are feasible on a limited scale, our research primarily addresses the late NISQ and early post-NISQ era, a stage subsequent to the attainment of quantum advantage. This advancement presupposes the non-simulability of state-of-the-art quantum applications on classical systems, rendering the simulation of even simplified ``toy'' examples negligible in value. This approach distinctly sets our work apart from prior models focused on earlier stages of QC development, where extensive simulation of large parts of the applications was still possible. We clarify that while simulations are not anticipated to become obsolete, their role in the development of practical applications will be significantly diminished, with their utility primarily confined to foundational research in algorithmic functions and mechanisms.

\paragraph{\textbf{Testing}}
Similar to classical software, quantum software should be tested as well \cite{zhao2020quantum}, although differences between both testing-cases are apparent \cite{garcia2023quantum}. Due to the underlying quantum mechanical effects, predicting quantum software behaviour is particularly difficult compared to classical software, which in turn makes it more prone to programming errors \cite{ying2012floyd}. The primary goal is to find all sorts of defects and verify the correct implementation of the intended functionalities before the solution is deployed. Research into bugs, assertions, testing, debugging and analysis of quantum software is therefore currently developing and should be closely monitored from all programmers \cite{zhao2020quantum}. The more extensive and critical the use of QC software becomes, the more important and far-reaching reliable test procedures become.  For solutions which were developed to become subsequently integrated into the existing IT landscape, testing must be carried out more extensively and in-depth, than if only a prototype implementation is aimed for \cite{soares2022effects}.

\paragraph{\textbf{Inference}}
After implementing, executing and testing the proposed quantum solutions, it is time to recap the concluded work. From experience we know that this point is often reached with several solutions, differing in the algorithms implemented, the heuristic methods applied, and / or the hardware used. This extra work that may have been done can now be leveraged. By integrating analysis tools and recording key performance indicators, it is possible to gain important insights into the various solutions. Thereby, we proceed in two directions. Firstly, through visualising results via figures and diagrams one can reveal improvement potential for the respective algorithm itself - a kind of individual view. Secondly, by comparing the different solutions, the best performing implementation can be identified - a kind of group view. Both views can help to further enhance the different solution paths individually, but also help to choose the best suitable solution at the second review-point. As this activity is one of the most important ones to boost performance, we would like to explain it using a very simplified example. \\ 
Given we have a VQE and a QAOA implementation, we can easily plot a convergence chart of their hybrid training, as in many SDKs the required data is stored per default. Inspecting the chart, one might exemplary identify that the training of the VQE algorithm stops before the gradients converge. The performance of the algorithm could now be increased by extending training. Additionally, by comparing the convergence charts of both algorithms, one might identify one algorithm which converges much faster then the other one. By assuming that the performance is identical in all other aspects, we are able to identify the best performing algorithm. This is of particular importance at the second review point, where the best algorithm is then selected. Even with this very simplified example we see, that through judiciously conducted inferences, a more refined and effective quantum solution can be realised, thereby justifying iterative cycles. The analysis of executions and the correct deduction of conclusions are therefore the pivotal point here.\\

As outlined, the stage of execution is highly iterative and is closely interlinked with the stages before. As the final result of this stage, an executable as well as efficient first solution is obtained, making this stage the most complex and timely extensive stage.

\subsection{Post-Processing}
The post-processing phase comprises activities that are carried out after the basic development of the solution algorithm. Its purpose is to further improve solution quality, again in an iterative fashion. 

\paragraph{\textbf{Error Handling \& Noise Mitigation}}
QC hardware is and will remain noisy in the near and medium term future. As a result, calculations are inaccurate and erroneous, which in turn reduces the usefulness of QC. Beside improving the hardware itself and reducing errors at its root, error-mitigation methods should be employed to alleviate the effects of noise. Two of the most known techniques are zero noise extrapolation (ZNE) \cite{kandala2019error, kim2023evidence, li2017efficient} and probabilistic error cancellation (PEC) \cite{temme2017error}. \\ Thereby, the ZNE, for example, deliberately varies the noise level inside the circuit in order to obtain results at different noise levels. Subsequently, the measurement results are extrapolated based on the results of the previous calculations to simulate the absence of noise. The method makes use of the property that the noise level can be artificially increased but cannot be reduced to a zero noise level. \\
The activity of error handling was intentionally put within the stage of post-processing. We argue that error handling can only be developed effectively once the application in its basic structure has been developed, as the characteristic of the the application influences the required error handling strategy. This activity is particularly unique compared to classic optimisation solutions.

\paragraph{\textbf{Post-Search}}
After the quantum algorithm has calculated an initial solution of the problem instance at hand, a post-search can be conducted. The quantum solution obtained is used as a starting point for a further classical optimisation or a second quantum optimisation run. In case of a second quantum optimisation, instead of a uniform super-position at the beginning of the algorithm, the pre-calculated solution, or at least parts of it, are used as initial starting point, with the intent to refine the solution \cite{borle2019post, perdomo2011study, alcazar2021geo}. 

\paragraph{\textbf{Scaling}}
Scaling can be applied to the input, as well as output data, to improve the performance of quantum machine learning models, particularly quantum reinforcement learning models \cite{skolik2022quantum, kolle2024study}. Input and output values are multiplied by a scalar, which itself is also part of the training process. This technique helps to align the scale of the input data with the scale of frequencies of the model function, as well as adjusts the output range, addressing the inherent limitations of PQCs in producing a restricted range of output values \cite{skolik2022quantum}. The empirical successes observed in environments like Cart Pole ~\cite{brockman2016openai} underscore the significance of these scaling techniques in QC applications and therefore qualifies such a technique as a potential activity in our development pipeline to further enhance performance. 

\paragraph{\textbf{Post-Selection}}
As final activity of the stage, the implementation of a post-selection mechanism is considered, with two categories being distinguishable: algorithmic-level post-selection and results-level post-selection. Within the algorithmic-level context, post-selection pertains to the selective processing of outcomes internal to the algorithm, exemplified by the HHL algorithm's approach \cite{harrow2009quantum}. The success of the computation is determined by the measurement of an ancilla qubit; a measurement outcome of $\ket{0}$ leads to the discarding of the calculation, whereas an outcome of $\ket{1}$ prompts the continuation of the computational process, thus effectively filtering results based on the ancilla qubit's state.\\
Conversely, the second category of post-selection is executed subsequent to the acquisition of the quantum algorithm's output, typically obtained as a probability distribution of bit-strings. This stage involves the selection of one or multiple results from the algorithm's output sample for final evaluation, which can be filtered by criteria such as cardinality \cite{alcazar2021geo}. The wide range of diverse post-selection methodologies is substantial, particularly given that such selections are often executed using straightforward criteria.

\subsection{Value Verification}

Once all stages have been completed, the second and last review-point is reached. Here we cover three aspects: choosing the best developed solution, the solution benchmarking against classical solution approaches, and the economic impact assessment of the newly developed quantum solution. \\
First of all, if several quantum solutions have been developed, we must select the optimal solution path. This then serves as a project outcome and will be benchmarked against classical, existing solutions. Solution benchmarking is central to determine the outcome of the project and its overall effectiveness. If no such current solution exists, one may consider to develop a classic solution here after all. A newly developed benchmarking solution does not have to be fully sophisticated, as its purpose is often just to give a rough orientation regarding solution quality. A simple off-the-shelf solution might be sufficient here, as we only want to estimate the relative performance for our own assessment and do not want to provide formal proof here. In order to subsequently determine the full performance of the solution, several metrics usually have to be taken into consideration, as quantum advantage can be reach with regards to several dimensions. The most common dimensions are time-to-solution and solution quality, whereby we strive for exponential speedup while maintaining or even improving solution quality. However, other advantages are also conceivable, such as lower costs to obtain the solution or lower CO2 emissions through a reduction in energy consumed by the quantum computer compared to a high-performance computing (HPC) centre. If QC provides a comparable solution to a HPC centre at a much lower cost, cost savings can be achieved if workload is shifted from classic HPC centre to quantum computers. Furthermore, financially far less interesting use-cases could now become relevant, as the development of a solution to these use-cases would now be worthwhile. \\
Finally, the economic impact is assessed. This aspect partly overlaps with the previously discussed aspects, but here we have a more strategic view, trying to capture the bigger picture. In addition to operational effects such as increasing efficiency in production or reduction of CO2 emissions through shorter transportation routes, QC may have a strategic impact on some industries and companies. With the adoption of quantum technologies, disruptive effects have to be considered. One example is the chemical industry, where new material might be discovered through quantum simulation. For leading companies of this industry, QC is not only a chance, but can also be seen as a threat. New players will evolve, trying to capture market shares by challenging market leaders. Established companies may thus find it necessary to invest in this technology to safeguard their market position from potential disruption. The question of whether the solution developed was a success and whether the company should continue to invest in further development and implementation has therefore also a strategic root. \\
As soon as both aspects have been considered in their diversity, an assessment can be made about the development and further procedure with regard to the underlying use-case. Although such a review may still seem relatively vague in times of a lack of quantum advantage, this review step will become increasingly important as QC matures.

\section{Conclusion}
In this paper, we proposed a general QC software development pipeline, particularly suited for optimisation problems. The proposed pipeline has five steps, two review points and numerous activities which structure the development process at hand. With this model, we try to consider the different perspectives of the various stakeholders involved in a quantum software project. We considered the most important technical facets, while also incorporating project management as well as strategic aspects into the model. Our objective was to foster mutual understanding and coordination across all involved stakeholders. However, also this research is not free of limitations. One limitation is the lack of practical application experience with this model under realistic conditions. Unfortunately, we have not yet had the opportunity to carry out a use-case or a complete QC software development project using the proposed pipeline. We are planning to use, evaluate and adapt the model to different application domains and concrete use-cases in those domains, as this is part of future work. An evaluation and adaptation to quantum annealing related development processes is also planned. Furthermore, the rapid development in the field of quantum software development and quantum algorithms means that new techniques and methods are constantly being developed that could be included in our pipeline as an activity. Therefore, until a certain level of standardisation in the development of quantum software solutions is reached, our pipeline should be revisited regularly, to check whether new solution techniques demand an update of our framework.

\section*{Acknowledgement}
This paper was partially funded by the German Federal Ministry of Education and Research through the funding program “quantum technologies -- from basic research to market” (contract number: 13N16196). J.S. acknowledges support from the German Federal Ministry for Economic Affairs and Climate Action through the funding program “Quantum Computing -- Applications for the industry” based on the allowance “Development of digital technologies” (contract number: 01MQ22008A).

\vspace{12pt}

\end{document}